\documentclass[pra,showpacs,twocolumn,nofootinbib]{revtex4}
\usepackage{graphicx}
\setkeys{Gin}{draft=false} 
\usepackage{bm,amsmath,textcomp}
\usepackage{dcolumn}
\usepackage{natbib}
\usepackage
{hyperref}

\newcommand{\etal}{{\it et al.}}

\newcommand{\Eref}[1]{Eq.~(\ref{#1})}
\newcommand{\tref}[1]{Table~\ref{#1}}

\newcommand{\peroxide}{{H$_2$O$_2$}}
\newcommand{\cm}{cm$^{-1}$}

\begin{document}

\title{Sensitivity of microwave transition
in \peroxide\ to variation of the electron-to-proton mass ratio}

\author{M. G. Kozlov}
\affiliation{Petersburg Nuclear Physics Institute, Gatchina 188300,
             Russia}
\date{
\today}

\begin{abstract}
Recent observation of several microwave transitions in \peroxide\ from the
interstellar medium [Bergman \etal\ Astron.~Astrophys., {\bf 531}, L8 (2011)]
raised interest to this molecule as yet another sensitive probe of the
tentative variation of the electron-to-proton mass ratio $\mu$. We estimate
sensitivity coefficients of the microwave transitions in \peroxide\ to
$\mu$-variation. The largest coefficient for 14.8 GHz transition is equal to
37, which is comparable to highest sensitivities in methanol and an order of
magnitude higher than sensitivity of the tunneling transition in ammonia.
\end{abstract}

 \maketitle


Molecules with an internal motion of large amplitude
recently attracted much attention as high sensitive probes of the tentative
variation of the electron-to-proton mass ratio $\mu=m_e/m_p$. In 2004
\citet{VKB04} pointed out that inversion transition in ammonia is very
sensitive to $\mu$-variation. Later, several groups used this transition to
place stringent limits on $\mu$-variation on a large timescale of the order of
10 Gyr \cite{FK07a,MFMH08,HMM09,Kan11}. At the same time ammonia spectra from
our Galaxy were used to study possible dependence of $\mu$ on the local mater
density \cite{LMK08,LLH10}. More recently the sensitivity to $\mu$-variation
was studied for other molecules with tunneling motion, including hydronium
(H$_3$O$^+$) \cite{KL11,KPR11}, methanol (CH$_3$OH) \cite{JXK11,LKR11,JKX11},
and methylamine (CH$_3$NH$_2$) \cite{IB11}. All these molecules are observed
in the interstellar medium (ISM) and potentially can be used as probes of
$\mu$-variation. Sensitivities of the tunneling transitions are two orders of
magnitude higher compared to optical transitions in molecular hydrogen, which
is traditionally used to study $\mu$-variation at high redshifts (see
\cite{WMM11} and references therein). Even higher sensitivities can be found
for certain mixed tunneling-rotational transitions. Recently methanol and
methylamine were detected at redshift $z=0.89$ \cite{MBG11}. First limit on
$\mu$-variation using methanol is reported in \cite{LKR11}.

Peroxide molecule (\peroxide) is one of the simplest molecules with large
amplitude tunneling mode. It is very well studied both theoretically and
experimentally. However, to the best of our knowledge, there is no analysis of
the sensitivity of the microwave transitions in this molecule to
$\mu$-variation. This is probably explained by the fact that \peroxide\ was
not observed in ISM. Situation has changed after the first observation of four
microwave lines of \peroxide\ from the molecular cloud core $\rho$ Oph A in
our Galaxy by \citet{BPL11}. In this communication we estimate sensitivities
of the microwave mixed tunneling-rotational transitions to $\mu$-variation.


In equilibrium geometry \peroxide\ is not flat; the angle $2\gamma$ between
two HOO planes is close to 113\textdegree. Two flat configurations correspond
to local maxima of potential energy; the potential barrier for \textit{trans}
configuration ($2\gamma=\pi$) is significantly lower, than for \textit{cis}
configuration ($\gamma=0$): $U_\pi\approx 400$ \cm\ and $U_0\approx 2500$ \cm.
To a first approximation one can neglect the tunneling through the higher
barrier. In this model peroxide is described by the slightly asymmetric oblate
top with inversion tunneling mode, similar to ammonia and hydronium. A more
accurate theory accounts for tunneling through both barriers \cite{Hou84}. In
this case torsion motion can be described as hindered rotation. For
sufficiently high angular quantum numbers $J$ and $K_A$ this internal motion
strongly interacts with overall rotation. According to \cite{MLF88} this
interaction becomes important for $K_A\ge 7$. Such levels lie very high and
can not be observed from ISM. Below 70K there are only levels with quantum
numbers $J \le 6$ and $K_A \le 2$. For such levels the simpler model is
sufficiently accurate and we will use it here.

When tunneling through both barriers is taken into account the ground torsion
state splits in four components designated by the quantum number $\tau=$ 1 --
4. Because \textit{trans} barrier is much lower than the \textit{cis} one, the
splittings between the pairs $\tau=1,2$ and $\tau=3,4$ is much larger then
splittings within the pairs. The latter is usually not resolved
experimentally. If we completely neglect second tunneling, we are left with
only two states, but for consistency they are labeled as $\tau=1$ and $\tau=3$
states \cite{Hou84,MLF88}.

Low energy effective Hamiltonian has the form (atomic units are used
throughout the paper):
\begin{align}\label{Heff1}
H_\mathrm{eff} &= E_\tau + H_\mathrm{rot}\,,
\end{align}
where $E_\tau$ is tunneling energy and $H_\mathrm{rot}$ is Hamiltonian of the
rigid asymmetric top, whose constants $A$, $B$, and $C$ weakly depend on the
quantum number $\tau$ (see \tref{tab1}).

\begin{table}[htb]
\caption{Parameters of the effective Hamiltonian \eqref{Heff1} in GHz from
Ref.\ \cite{MLF88}.}
  \label{tab1}
\begin{tabular}{cdddd}
\hline\hline
 \multicolumn{1}{c}{$\tau$}
 &\multicolumn{1}{c}{$E_\tau$}
 &\multicolumn{1}{c}{$A_\tau$}
 &\multicolumn{1}{c}{$B_\tau$}
 &\multicolumn{1}{c}{$C_\tau$}
\\
\hline
 1   &  0.0  &301.873 &26.193  &25.120 \\
 3   &342.885&301.585 &26.142  &25.201 \\
\hline\hline
\end{tabular}
\end{table}

Dimensionless sensitivity coefficient $Q_\mu$ of the transition $\omega$ to
$\mu$-variation is defined so that:
 \begin{equation}\label{Q1}
 \frac{\Delta \omega}{\omega} = Q_\mu\frac{\Delta\mu}{\mu}\,.
 \end{equation}
In order to find these coefficients for transitions described by Hamiltonian
\eqref{Heff1} we need to know $\mu$-dependence of the parameters from
\tref{tab1}. Rotational parameters $A$, $B$, and $C$ scale linearly with
$\mu$. Thus, purely rotational transition has sensitivity
$Q_{\mu,\mathrm{rot}}=1$. Sensitivity of the tunneling energy $E_\tau$ to
$\mu$-variation can be estimated with the help of semiclassical
Wentzel-Kramers-Brillouin approximation. Following \cite{LL77}, we can write:
\begin{equation}\label{inv1}
 E_\tau = \frac{2E_0}{\pi}\,\mathrm{e}^{-S},
\end{equation}
where $S$ is the action over classically forbidden region and $E_0$ is the
zero point energy
 for the inversion mode.
 Expression \eqref{inv1} gives the
following sensitivity to $\mu$-variation:
\begin{equation}\label{inv2}
 Q_{\mu,\tau}= \frac{S+1}{2}
 +\frac{S\,E_0}{2(U_\mathrm{max}-E_0)}\,,
\end{equation}
where $U_\mathrm{max}$ is the barrier hight.
Numerical factor in the second fraction depends on the barrier shape. For the
triangular and square barriers it is 2 times smaller and 1.5 times larger
respectively \cite{FK07a}. The factor $\tfrac12$ in \Eref{inv2} corresponds to
the parabolic barrier \cite{KPR11}.

The potential for the tunneling coordinate $\gamma$ was found by \citet{KCH98}
in a form of Fourier expansion with $U_\mathrm{max}= U_\pi=387$~\cm. Numerical
solution of the one dimensional Schr\"odinger equation for this potential
gives $E_0=169$~\cm\, $E_\tau=13.1$~\cm\ and $Q_{\mu,\tau}=2.44$. Using this
value for $E_0$ and tunneling energy from \tref{tab1} we can find $S$ from
\Eref{inv1} and $Q_{\mu,\tau}$ from \Eref{inv2}:
\begin{equation}\label{inv3}
 S= 2.28\,,\qquad Q_{\mu,\tau}= 2.54\,.
\end{equation}
If we scale potential from \cite{KCH98} to fit experimental tunneling
frequency from \tref{tab1}, the numerical solution gives $Q_{\mu,\tau}=2.56$.
All three values for $Q_{\mu,\tau}$ agree to 5\%.
The semiclassical value corresponds to the experimental tunneling frequency
and is less sensitive to the details of the potential shape. Therefore, we use
it in our calculations and assign it 5\% uncertainty.

We see, that tunneling energy is 2.5 times more sensitive to $\mu$-variation
than rotational energy. Sensitivity of the mixed tunneling-rotational
transitions
$ \omega = E_\tau \pm \omega_\mathrm{rot},$ is a weighted average of the
tunneling and rotational contributions \cite{KLL10}:
\begin{equation}\label{mixed}
 Q_\mu= \frac{E_\tau}{\omega} Q_{\mu,\tau}
 \pm \frac{\omega_\mathrm{rot}}{\omega} Q_{\mu,\mathrm{rot}}\,.
\end{equation}
This sensitivity is further enhanced for the frequencies $|\omega| \ll
E_\tau$.

Results of the numerical calculations with Hamiltonian \eqref{Heff1} are given
in \tref{tab_Q}. In this table we list six transitions in the range 200 -- 700
GHz, which were studied in Ref.\ \cite{BPL11} (four of them were observed and
other two were not) and transitions from the JPL database \cite{JPL_Catalog}
with frequencies below 100 GHz.

\begin{table}[bt!]
\caption{Numerical calculation of the $Q$-factors for low frequency mixed
transitions in peroxide using Hamiltonian \eqref{Heff1} and \Eref{inv3}.
Experimental frequencies are taken from JPL Catalogue \cite{JPL_Catalog}.
$E_\mathrm{up}$ is upper state energy in Kelvin.}
  \label{tab_Q}
\begin{tabular}{c@{ -- }ccrdd}
\hline\hline
 \multicolumn{2}{c}{$J_{K_A,K_C}(\tau)$}
 &\multicolumn{1}{c}{$E_\mathrm{up}$}
 &\multicolumn{2}{c}{$\omega$ (MHz)}
 &\multicolumn{1}{c}{$Q_\mu$}
\\
 \multicolumn{1}{c}{upper}
 &\multicolumn{1}{c}{lower}
 &\multicolumn{1}{c}{(K)}
 &\multicolumn{1}{c}{theory}
 &\multicolumn{1}{c}{exper.}
 &
\\
\hline
\multicolumn{6}{c}{Transitions below 100 GHz}\\
$ 0_{0,0}(3) $&$ 1_{1,0}(1) $& 17& 14818.8    &    14829.1  &   +36.5(2.9) \\
$ 2_{1,1}(1) $&$ 1_{0,1}(3) $& 21& 37537.0    &    37518.28 &   -13.0(1.2) \\
$ 1_{0,1}(3) $&$ 1_{1,1}(1) $& 19& 67234.5    &    67245.7  &    +8.8(6) \\
$ 2_{0,2}(3) $&$ 2_{1,2}(1) $& 24& 68365.3    &    68385.0  &    +8.7(6) \\
$ 3_{0,3}(3) $&$ 3_{1,3}(1) $& 31& 70057.4    &    70090.2  &    +8.5(6) \\
$ 4_{0,4}(3) $&$ 4_{1,4}(1) $& 41& 72306.0    &    72356.4  &    +8.3(6) \\
$ 5_{0,5}(3) $&$ 5_{1,5}(1) $& 53& 75104.6    &    75177.4  &    +8.0(6) \\
$ 6_{0,6}(3) $&$ 6_{1,6}(1) $& 68& 78444.7    &    78545.4  &    +7.7(6) \\
$ 3_{1,2}(1) $&$ 2_{0,2}(3) $& 28& 90399.8    &    90365.51 &    -4.8(5) \\[1mm]
\multicolumn{6}{c}{Transitions observed from ISM in Ref.\ \cite{BPL11}}\\
$ 3_{0,3}(3) $&$ 2_{1,1}(1) $& 31&219163.2    &   219166.9  &    +3.4(2) \\
$ 6_{1,5}(1) $&$ 5_{0,5}(3) $& 66&252063.6    &   251914.68 &    -1.1(2) \\
$ 4_{0,4}(3) $&$ 3_{1,2}(1) $& 41&268963.7    &   268961.2  &    +3.0(2) \\
$ 5_{0,5}(3) $&$ 4_{1,3}(1) $& 53&318237.7    &   318222.5  &    +2.7(1) \\[1mm]
\multicolumn{6}{c}{Transitions attempted to observe in Ref.\ \cite{BPL11}}\\
$ 5_{1,4}(3) $&$ 6_{0,6}(1) $& 67&318635.6    &   318712.1  &    +2.7(1) \\
$ 1_{1,0}(3) $&$ 0_{0,0}(1) $& 32&670611.9    &   670595.8  &    +1.8(1) \\
\hline\hline
\end{tabular}
\end{table}

Note that pure rotational transitions with $\Delta \tau=0$ for peroxide are
not observed. Because of that all sensitivities in \tref{tab_Q} significantly
deviate from unity. As expected, lower frequency transitions have higher
sensitivities. Transitions with the frequency $|\omega|<E_\tau$ fall in two
categories. For transitions $\tau=3 \rightarrow \tau=1$ the tunneling energy
is larger than rotational energy and $|\omega|=E_\tau-\omega_\mathrm{rot}$.
For such transitions $Q$-factors are positive. For transitions $\tau=1
\rightarrow \tau=3$, $|\omega|=\omega_\mathrm{rot}-E_\tau$ and $Q$-factors are
negative.

Let us discuss the accuracy of our estimates. Calculated frequencies agree
with experiment to 0.1\%, or better in spite of the simplicity of Hamiltonian
\eqref{Heff1}. That means that centrifugal corrections to the rotational
energy are indeed unimportant for the transitions with low rotational quantum
numbers considered here. Even in this case the scaling of the rotational
energy with $\mu$ is not exactly linear as rotational constants of the
effective Hamiltonian are averaged over the vibrational wave functions of the
states $\tau=1,3$. Respective corrections to the rotational sensitivity
coefficients are of the order of 1\% -- 2\% \cite{LKR11}. Note that
$\tau$-dependence of the rotational constants can be considered as centrifugal
corrections to the tunneling energy \cite{KPR11}. According to Refs.\
\cite{KPR11,LKR11} the accuracy of the semiclassical expressions for tunneling
(\ref{inv1}, \ref{inv2}) is comparable, about 3\%. Additional uncertainty is
associated with the zero point energy $E_0$, which is not directly observable
and we take it from calculation with potential \cite{KCH98}. This potential
was used in a number of later papers, including \cite{LG03,CSB11}. We estimate
the accuracy of the tunneling sensitivity coefficient \eqref{inv3} to be 5\%,
which agrees with numerical estimates made above. Uncertainties in rotational
and tunneling sensitivities result in the final uncertainties for mixed
transitions given in \tref{tab_Q}. These uncertainties are typically about
10\%. Such accuracy is sufficient for the analysis of the astrophysical
spectra.

To summarize, we have estimated sensitivity coefficients for the low frequency
mixed tunneling-rotational transitions in \peroxide. Transitions below 100 GHz
appear to be highly sensitive to $\mu$-variation with sensitivity coefficients
of both signs.
Maximal relative sensitivity is found for transitions 14.8 and 37.5 GHz, where
$\Delta Q_\mu=+36.5-(-13.0)\approx 50$.
This is comparable to the highest sensitivities in methanol \cite{JXK11,LKR11}
and an order of magnitude larger than for ammonia \cite{VKB04,FK07a}. The
lines recently observed by \citet{BPL11} have frequencies above 200 GHz and
significantly lower sensitivities. Three of the observed transitions have
sensitivities, which are close to each other. However, the sensitivity of the
fourth observed transition is significantly different. Transitions 219 GHz and
252 GHz have relative sensitivity $\Delta Q_\mu=4.5$. This is close to the
sensitivity of the ammonia method, where $\Delta Q_\mu=3.5$, but with the
advantage that both lines belong to one species. This eliminates very
important source of systematic errors caused by the difference in spatial and
velocity distributions of species \cite{LLH10}. We conclude, that microwave
transitions in \peroxide\ potentially can be used for $\mu$-variation search
in astrophysics. Note, that high sensitivity transitions correspond to low
rotational quantum numbers $J$. This makes peroxide a potential candidate for
laboratory tests on $\mu$-variation using molecular beam technique
\cite{JB11}.\\[2mm]


I am grateful to S.\ A.\ Levshakov, A.\ V.\ Lapinov, P.\ Jansen, and H.\ L.\
Bethlem for helpful discussions. This work was supported by RFBR grants
11-02-00943 and 11-02-12284.


\end{document}